\documentclass[
    ,final            
  ]
  {aipproc}

\layoutstyle{6x9}

\usepackage{amsmath}

\usepackage{amssymb}

\usepackage{array}

\usepackage{a4wide}

\usepackage{eucal}

\usepackage{scalefnt}

\usepackage[english]{babel}

\usepackage{verbatim}

\usepackage{psfrag}

\usepackage{epsfig}

\newlength{\textlength}
\newlength{\overlinelength}
\newcommand{\ovl}[2][.55]{\settowidth{\textlength}{$#2$}
  \setlength{\overlinelength}{0.1pt}
  \addtolength{\overlinelength}{0.75\textlength}
  \makebox[\textlength][s]{$#2$} \hspace{-.55\textlength}
  \hspace{-\overlinelength}\hspace{#1\overlinelength}
  \overline{\makebox[\overlinelength][s]{\vphantom{$#2$}}}
  \hspace{-#1\overlinelength}\hspace{.55\textlength}}

\begin{document}

\title{Grand Unified Flavour Physics}


\author{Susanne Westhoff}{
  address={Institut f\"ur Theoretische Teilchenphysik, Universit\"at Karlsruhe, D-76128 Karlsruhe, Germany}
}

\begin{abstract}
We probe the unification of down quarks and leptons in a supersymmetric $\text{SO}(10)$ GUT. The large atmospheric neutrino mixing angle induces $b_R-s_R$ transitions, which can account for the sizeable $CP$ phase $\phi_s$ measured in $B_s-\ovl{B}_s$ mixing. Corrections to down-quark-lepton unification from higher-dimensional Yukawa terms translate neutrino mixing also into $s_R-d_R$ and $b_R-d_R$ currents. We find the flavour structure of Yukawa corrections to be strongly constrained by $\epsilon_K$.
\end{abstract}

\maketitle

The embedding of the Standard Model into a larger symmetry group implies relations among Standard-Model parameters, which can be tested by experiment. In the framework of supersymmetric Grand Unified Theories, the unification of gauge couplings has passed such tests excellently. Flavour physics probes the unification of Yukawa couplings in low-energy observables. The unification of down (s)quarks and (s)leptons can transfer large mixing in the lepton sector into flavour-changing neutral currents among down squarks. In particular, the atmospheric neutrino mixing angle leaves imprints of Yukawa unification in $B_s$ obser\-vables. For light fermions, the mass relations between down quarks and leptons need to be corrected. To this end, one introduces higher-dimensional Yukawa terms that a priori imply large mixing effects also in $K$ and $B_d$ physics observables. The precise knowledge of these observables allows to constrain the flavour structure of such Yukawa corrections. We show that corrections to down-quark-lepton unification must be basically aligned with the initial Yukawa couplings in flavour space.
\section{GUT relations between down quarks and leptons}
The unification of down quarks and leptons is a feature of $\text{SU}(5)$ symmetry. Explicitly, the $\text{SU}(2)$-singlet down-type quarks $d^c$ are combined with the lepton doublet $L=(\ell,\nu_{\ell})$ in a $\ovl{5}$ representation of $\text{SU}(5)$,
\begin{align}
\ovl{5}=(d_1^c,\,d_2^c,\,d_3^c; \ell,\,\nu_{\ell})\,,
\end{align}
where $1,2,3$ are colour indices and a superscript $c$ denotes charge conjugation. The quark doublet $Q$, as well as the quark and lepton singlets $u^c$ and $\ell^c$, are embedded into a $10$. The Yukawa sector comprises fermion couplings to the Higgs representations $5_H\supset H_u$, $\ovl{5}_H\supset H_d$, which contain the Higgs doublets  $H_u$ and $H_d$ of a two-Higgs-doublet model,
\begin{align}
W_Y=10_i\,Y_5^{ij}\,10_j\,5_H+10_i\,Y_{\bar{5}}^{ij}\,\ovl{5}_{j}\,\ovl{5}_{H}\,.
\end{align}
The first term gives masses to up quarks, whereas the second term generates masses for both down quarks and leptons. Thereby the Standard-Model Yukawa couplings $Y_d$ and $Y_{\ell}$ are unified at the scale of gauge coupling unification $M_{\text{GUT}}$,
\begin{align}\label{dluni}
Y_d=Y_{\ell}^{\top}=Y_{\bar{5}}\,.
\end{align}
By evolving this relation down to the electroweak scale $M_Z$ and comparing with the measured down-quark and lepton masses, one learns that experimentally down-quark-lepton unification holds only for the bottom and tau couplings. Corrections for the mass relations between light fermions can be provided by adding terms of mass dimension five to the Yukawa sector \cite{Ellis1979},
\begin{align}\label{su5dim5}
W_Y^{d,\ell}=10_i\,Y_{\bar{5}}^{ij}\,\ovl{5}_{j}\,\ovl{5}_{H}+10_i\,\frac{24_H}{M_{\text{Pl}}}\,Y_{\sigma}^{ij}\,\ovl{5}_{j}\,\ovl{5}_{H}\,.
\end{align}
If the adjoint Higgs representation $24_H$ acquires a vacuum expectation value proportional to $\text{SU}(5)$ hypercharge, 
$\langle 24_H\rangle=\sigma\,\text{diag}(2,2,2;-3,-3)$, relation (\ref{dluni}) is modified below the GUT scale to
\begin{align}\label{Yukawacorr}
Y_d=Y_{\ell}^{\top}+5\frac{\sigma}{M_{\text{Pl}}}Y_{\sigma}\,.
\end{align}
Due to the suppression by $\sigma/M_{\text{Pl}}=\mathcal{O}(10^{-3})$, the Yukawa corrections $Y_{\sigma}$ affect only the light fermions, while preserving the successful bottom-tau Yukawa unification.
\section{Neutrino mixing in down-squark interactions}
Yukawa corrections from dimension-five terms can be implemented straight into $\text{SO}(10)$, where all fermions of one generation, plus a heavy right-handed neutrino $N$, are unified in one spinor representation $16=1\oplus10\oplus\ovl{5}=(N,\,(Q,\,u^c,\,\ell^c),\,(d^c,\,L))$.
We examine a specific supersymmetric $\text{SO}(10)$ model proposed by Chang, Masiero, and Murayama (CMM) \cite{CMM2003}. The Yukawa sector is given by
\begin{align}\label{wcmm}
  W_{\text{CMM}} & = \text{16}_i\, \widehat{Y}^{ii}_{10}\, \text{16}_i\, \text{10}_H
  + \text{16}_i\, (V_{q}^{\ast}\,\widehat{Y}_{45}\,V_{\ell})^{ij}\, \text{16}_j\,
  \frac{\text{45}_H\, \text{10}'_H}{M_\text{Pl}} + \text{16}_i\,
  \widehat{Y}^{ii}_{16}\, \text{16}_i \frac{\ovl{\text{16}}_H
    \ovl{\text{16}}_H}{2M_\text{Pl}}\,,
\end{align}
where the three terms generate masses for up quarks, down quarks and leptons, and right-handed neutrinos, respectively. Under the assumption that the up-quark and neutrino Yukawa couplings $Y_{10}$ and $Y_{16}$ are simultaneously diagonal, flavour mixing is fully contained in the down-quark-lepton coupling $Y_{45}$. In the fermion mass eigenbasis, the mixing matrices $V_q$ and $V_{\ell}$ are identified with $V_{\text{CKM}}$ and $V_{\text{PMNS}}$ up to phases. Due to down-quark-lepton unification, $V_{\text{PMNS}}$ rotates both leptons and right-handed down quarks. The large atmospheric neutrino mixing angle $\theta_{23}\simeq 45^{\circ}$ thereby translates into significant $b_R-s_R$ transitions.

As in $\text{SU}(5)$, corrections to Yukawa unification for light fermions arise from the $\text{SU}(5)$-breaking vacuum expectation value $\langle 24_H\rangle$, since the $45_H$ of $\text{SO}(10)$ contains the $24_H$ of $\text{SU}(5)$. They generate additional rotations of down quarks with respect to leptons, parametrized by a mixing angle $\theta$. Up to phases, the rotation matrix of right-handed down quarks is thus given by
\begin{align}
R_d=(UV_{\ell})^{\top},\qquad\quad U=\begin{pmatrix}
                                     \cos\theta & -\sin\theta & 0\\
				     \sin\theta & \cos\theta & 0\\
				     0 & 0 & 1
                                    \end{pmatrix}.
\end{align}
As a consequence of this additional rotation, the atmospheric neutrino mixing angle also appears in $s_R-d_R$ and $b_R-d_R$ transitions. In a supersymmetric framework, these effects are visible in the couplings of down squarks if the mass spectrum of scalar superpartners is not degenerate. We assume flavour-blind supersymmetry breaking at the Planck scale, which corresponds to universal soft masses $m_0^2$ and trilinear couplings $A_0$ at $M_{\text{Pl}}$. Effects of the large top Yukawa coupling in the $\text{SO}(10)$ renormalization group evolution induce a mass splitting $\Delta_{\tilde{d}}=\mathcal{O}(0.5)$ in the down-squark soft mass matrix (here in the up-quark mass eigenbasis),
\begin{align}
M_{\text{Pl}}:\ (M_{\tilde{d}}^2)^U=m_0^2\cdot\begin{pmatrix}
1 & 0 & 0 \\
0 & 1 & 0 \\
0 & 0 & 1
\end{pmatrix}\quad\stackrel{\text{RGE}}{\longrightarrow}\quad M_Z:\ (M_{\tilde{d}}^2)^U=m_{\tilde{d}}^2\cdot\begin{pmatrix}
1 & 0 & 0 \\
0 & 1 & 0 \\
0 & 0 & 1-\Delta_{\tilde{d}}
\end{pmatrix}.
\end{align}
In the super-CKM basis, where the scalar superpartners are simultaneously rotated with the fermions, the mass matrix of singlet down squarks reads
\begin{align}\label{mdsCKM}
(M^2_{\tilde{d}})^{\text{sCKM}} & = R_d^\dagger\,(M^2_{\tilde{d}})^U\, R_d\nonumber\\
& =m_{\tilde{d}}^2\cdot
\left(\begin{array}{ccc}
1-\sin^2\theta\,\Delta_{\tilde{d}}/2 &
\sin(2\theta)\,e^{-i\phi_K}\,\Delta_{\tilde{d}}/4 &
\sin\theta\,e^{-i\phi_{B_d}}\,\Delta_{\tilde{d}}/2 \\
\hspace*{-0.2cm}\sin(2\theta)\,e^{i\phi_K}\,\Delta_{\tilde{d}}/4 &
1-\cos^2\theta\,\Delta_{\tilde{d}}/2 &
-\cos\theta\,e^{-i\phi_{B_s}}\,\Delta_{\tilde{d}}/2 \\
\sin\theta\,e^{i\phi_{B_d}}\,\Delta_{\tilde{d}}/2 &
-\cos\theta\,e^{i\phi_{B_s}}\,\Delta_{\tilde{d}}/2 &
1-\Delta_{\tilde{d}}/2
\end{array}\hspace*{-0.2cm}\right)\,,
\end{align}
assuming tri-bi-maximal lepton mixing. The off-diagonal elements induce effects of atmospheric neutrino mixing in $K$, $B_d$, and $B_s$ physics observables. The phases $\phi_K$, $\phi_{B_d}$, and $\phi_{B_s}=\phi_{B_d}-\phi_K$ are new sources of $CP$ violation. We will concentrate on the characteristic signatures from down-quark-squark gluino box diagrams in meson mixing.

\begin{figure}[t]
\begin{minipage}{0.5\textwidth}
\hspace*{0.5cm}\scalebox{0.7}{\epsfig{file=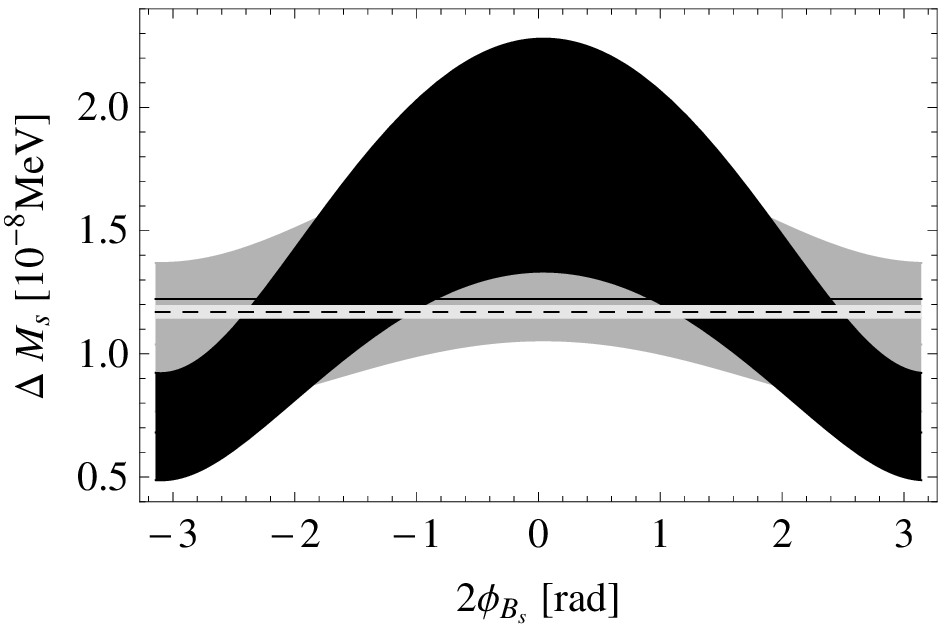}}
\end{minipage}
\begin{minipage}{0.5\textwidth}
\hspace*{0.5cm}\scalebox{0.7}{\epsfig{file=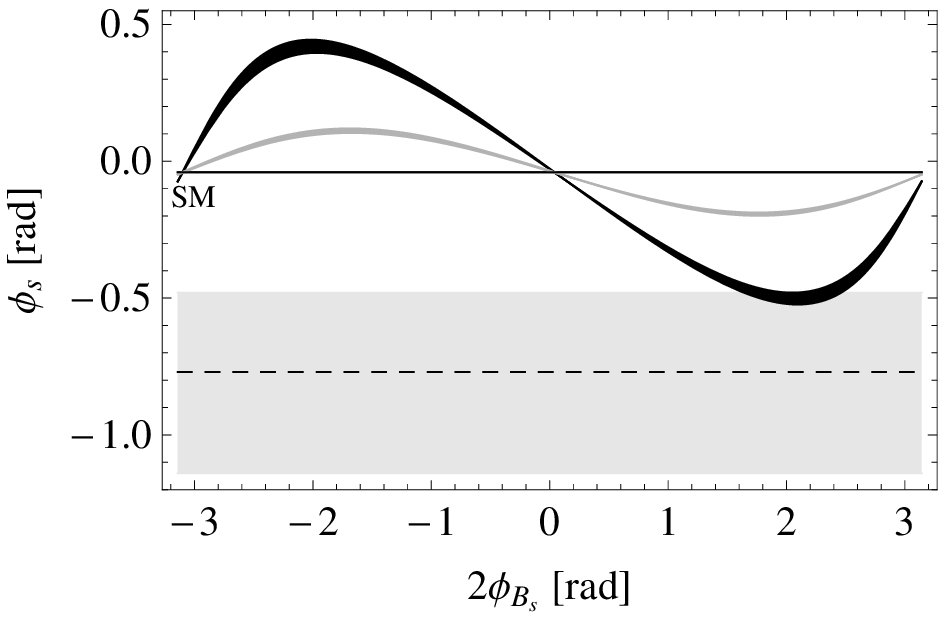}}
\end{minipage}
\caption{Effects of large atmospheric neutrino mixing in $B_s-\ovl{B}_s$ mixing for fixed inputs $m_{\tilde{d}}=2\,\text{TeV}$, $a_d/m_{\tilde{d}}=1.8,\,\text{arg}(\mu)=0,\,\tan\beta=5$, and $m_{\tilde{g}}=400\,\text{GeV}$ (black curve), $700\,\text{GeV}$ (gray curve). Left: $\Delta M_s$, gray band: experimental three-sigma range. Right: $\phi_s=\arg(-M_{12}/\Gamma_{12})$, gray band: experimental one-sigma range. Standard-Model values are indicated by a black line.}\label{fig:phibs}
\end{figure}

For $\theta=0$, corresponding to vanishing Yukawa corrections, imprints of the atmospheric neutrino mixing angle are confined to $B_s$ observables. In $B_s-\ovl{B}_s$ mixing, both the mass difference $\Delta M_s$ and the $CP$-violating phase $\phi_s$ receive significant contributions. The SUSY spectrum in the CMM model is determined by five input parameters at $M_Z$, which are the gluino mass $m_{\tilde{g}}$, the universal down-squark mass $m_{\tilde{d}}$, the ratio of down-quark trilinear and Yukawa couplings $a_d=(A_d)^{11}/(Y_d)^{11}$, the phase of the higgsino mass parameter $\text{arg}(\mu)$, and the ratio of the Higgs-doublet vacuum expectation values $\tan\beta$. The renormalization group links these inputs via the universality conditions at $M_{\text{Pl}}$ to the entire SUSY spectrum at low scales. In Fig.~\ref{fig:phibs} left, we show the effects of large atmospheric neutrino mixing on $\Delta M_s$ for two typical sets of input parameters distinguished by $m_{\tilde{g}}$. For gluino masses $m_{\tilde{g}}\simeq 400\,\text{GeV}$, the CMM phase is constrained from $\Delta M_s$ to $1.2\lesssim |2\phi_{B_s}|\lesssim 2.4$. Comparing with Fig.~\ref{fig:phibs} right, one observes that within this parameter region the $CP$ phase in $B_s-\ovl{B}_s$ mixing amounts to $\phi_s=-0.5$ (contrarily to the tiny $\phi_s=-0.04$ in the SM). The CMM model can thus account for the measured phase $\phi_s^{\text{exp}}=-0.77^{\,+\,0.29}_{\,-\,0.37}$ or $-2.36^{\,+\,0.37}_{\,-\,0.29}$ \cite{HFAG2008}, while fulfilling the constraints from $\Delta M_s$.
\section{Constraints on Yukawa corrections from $\epsilon_K$}
The magnitude of neutrino mixing effects in $K-\ovl{K}$ and $B_d-\ovl{B}_d$ mixing from Yukawa corrections $Y_{\sigma}$ is parametrized by the rotation angle $\theta$, which a priori can be sizeable. The $CP$-violating observable $\epsilon_K$ in the kaon system, however, sets strong constraints on $\theta$. $\epsilon_K$ is very well measured, theoretically rather clean, and extremely sensitive to new-physics contributions. One derives the upper bound $\theta^{\text{max}}\simeq 1^{\circ}$ if the phase $\phi_K$ is different from $\phi_K=0,\pi/2$ \cite{TWW2009}. Excluding any miraculous cancellations, this bound allows to specify the flavour structure of Yukawa corrections:\vspace{0.5cm}
\begin{tabular}{c}
\hspace{1.3cm}\framebox{\rule{0cm}{0.3cm}The Yukawa corrections $Y_{\sigma}$ have to be aligned with $Y_d$ and $Y_{\ell}$, see Eq.~(\ref{Yukawacorr}).}
\end{tabular}\vspace{0.5cm}\\
In other words, corrections to down-quark-lepton unification for light fermions cannot introduce new flavour structures with respect to the initial unified Yukawa couplings if the constraint from $\epsilon_K$ holds. Effects of large neutrino mixing in less sensitive $K$ and $B_d$ observables are consequently negligibly tiny. In the case of vanishing effects in $\epsilon_K$, $\theta$ can still be restrained from $B-\ovl{B}$ mixing observables to $\theta^{\text{max}}\simeq 20^{\circ}$ \cite{TWW2009}. 

This finding is of general interest for GUT model building: Once down-quark-lepton Yukawa unification of light fermions is corrected, $\epsilon_K$ strongly constrains the penetration of large atmospheric neutrino mixing into $s_R-d_R$ and $b_R-d_R$ currents. In the CMM model, effects are induced by the large down-squark mass splitting $\Delta_{\tilde{d}}$ due to $\text{SO}(10)$ running. There might be other sources that break the squark mass universality, like down-squark couplings to heavy right-handed neutrinos in supersymmetric $\text{SU}(5)$ models \cite{Hall1986}. Whenever such a mass splitting occurs, the flavour structure of corrections to down-quark-lepton Yukawa unification is restricted to be aligned with the initial couplings.
\begin{center}
$\ast\ast\ast$
\end{center}
I thank my collaborators St\'ephanie Trine and S\"oren Wiesenfeldt. This work is supported by the DFG-SFB/TR9 and by the DFG Graduiertenkolleg ``Hoch\-energiephysik und Astro\-teilchenphysik''.
\bibliographystyle{aipproc1}

\begin{thebibliography}{9}

\bibitem{Ellis1979}
J.~Ellis and M.~Gaillard, \emph{Phys.~Lett.} {\bf B88} (1979) 315.

\bibitem{CMM2003}
D.~Chang, A.~Masiero, and H.~Murayama, \emph{Phys.~Rev.} {\bf D67} (2003) 075013, arXiv:hep-ph/0205111.

\bibitem{HFAG2008}
Heavy Flavor Averaging Group, E.~Barberio et al., arXiv:0808.1297. Average of the D0 and CDF measurements.

\bibitem{TWW2009}
S.~Trine, S.~Westhoff, and S.~Wiesenfeldt, \emph{JHEP} {\bf 08} (2009) 002, arXiv:0904.0378.

\bibitem{Hall1986}
L.~Hall, A.~Kostelecky, and S.~Raby, \emph{Nucl.~Phys.} {\bf B267} (1986) 415.

\end{thebibliography}


\end{document}